\documentclass
[superscriptaddress,secnumarabic,amssymb,amsmath,nobibnotes,aps,prd,showkeys,showpacs,nofootinbib,onecolumn,eqsecnum]{revtex4}%
\usepackage{graphics}
\usepackage{graphicx}
\usepackage{color}
\usepackage{epsf}
\usepackage{bm}
\usepackage{amsmath,amssymb,amsfonts,mathrsfs,amsthm}
\usepackage{latexsym}
\usepackage{enumerate}
\usepackage{comment}
\usepackage{hyperref}
\usepackage{amsmath}
\usepackage{amsfonts}
\usepackage{amssymb}%
\providecommand{\U}[1]{\protect\rule{.1in}{.1in}}

\newcommand{\be}{\begin{equation}}
\newcommand{\ee}{\end{equation}}
\newcommand{\bea}{\begin{eqnarray}}
\newcommand{\ea}{\end{eqnarray}}
\newcommand{\ben}{\begin{equation*}}
\newcommand{\een}{\end{equation*}}
\newcommand{\bean}{\begin{eqnarray*}}
\newcommand{\eean}{\end{eqnarray*}}
\def\bal#1\eal{\begin{align}#1\end{align}}

\newcommand{\mincir}{\raise
-3.truept\hbox{\rlap{\hbox{$\sim$}}\raise4.truept\hbox{$<$}\ }}
\newcommand{\magcir}{\raise
-3.truept\hbox{\rlap{\hbox{$\sim$}}\raise4.truept\hbox{$>$}\ }}

\begin{document}
\title{Exact solutions of Einstein-\ae ther gravity in Bianchi Type V Cosmology}
\author{M. Roumeliotis}
\email{microum@phys.uoa.gr}
\affiliation{Nuclear and Particle Physics section, Physics Department, University of
Athens, 15771 Athens, Greece}
\author{A. Paliathanasis}
\email{anpaliat@phys.uoa.gr}
\affiliation{Institute of Systems Science, Durban University of Technology, POB 1334 Durban
4000, South Africa.}
\author{Petros A. Terzis}
\email{pterzis@phys.uoa.gr}
\affiliation{Nuclear and Particle Physics section, Physics Department, University of
Athens, 15771 Athens, Greece}
\author{T. Christodoulakis}
\email{tchris@phys.uoa.gr}
\affiliation{Nuclear and Particle Physics section, Physics Department, University of
Athens, 15771 Athens, Greece}

\begin{abstract}
We present the solution space of the field equations in the Einstein-\ae ther
theory for the case of a vacuum Bianchi Type $V$ space-time. We also
find that there are portions of the initial parameters space for which no
solution is admitted by the reduced equations. Whenever solutions do exist,
their physical interpretation is examined through the behavior of Ricci and/or
Kretsmann scalar, as well as with the identification of the effective energy
momentum tensor in terms of a perfect fluid. There are cases in which  no
singularities appear and others where the effective fluid is isotropic.

\end{abstract}
\maketitle
\date{\today}

\section{Introduction}

Bianchi spacetimes contain various important cosmological models that have
been used to describe the anisotropies of primordial universe and the
evolution to an isotropic universe as observed in the present epoch.
\cite{Mis69,jacobs2,collins,JB1}. In Bianchi models the spacetime manifold is
foliated along the time axis, with three dimensional homogeneous hypersurfaces
\cite{bk1}. The corresponding geometries are thus spatially homogeneous, which
means that the physical variables depend on the time variable only, reducing
the Einstein and other governing equations to coupled ordinary differential
equations. In total there are nine Bianchi models which are defined from the
possible nine different three-dimensional algebras constructed from the isometries of the metric tensor, see for example \cite{bk1,Ter}.

In the context of GR, the field equations for Bianchi spacetimes admit exact
solutions, for instance see
\cite{bi,bii,biia,biii,biiia,biv,bv,bvi,bvii,bviii,bix} and references
therein. In the case where a matter source is included in Einstein's GR, such
as electromagnetic fluid, ideal gas and scalar field, some exact solutions have
been determined previously in
\cite{sol1,sol2,sol3,sol4,sol5,sol6,sol7,sol8,sol9,sol10}.

In this work we are interested in the exact solutions of field equations for
Bianchi V spacetime in Einstein-aether gravity \cite{DJ,DJ2}. This is a
Lorentz violating gravitational theory, because a preferred frame is introduced
through the aether field present in the gravitational action integral
\cite{ea1}. Mathematically, the aether field is described by a unit
time-like vector field, with its kinematic quantities coupled to gravity in
the Einstein-Hilbert action. By definition Einstein-\ae ther gravity is a
second-order theory however, because of the existence of nonlinear terms in
the aether field, there are very few exact solutions. For applications of
Einstein-aether gravity in cosmological studies we refer the reader in
\cite{eap1,eap2,eap3,eap4,eap5}.

Recently, in \cite{roum} exact solutions where found for the case of
Friedmann--Lema\^{\i}tre--Robertson--Walker (FLRW) and LRS Bianchi III
spacetimes in Einstein-aether theory. It was there proved that exact solutions
exist only for specific ranges of values of the free coupling parameters for
the aether field. These exact solutions have been applied in \cite{an1} for the
determination of inhomogeneous cosmological solutions in Einstein-aether
theory. Other studies which can be found in the literature on anisotropic
Einstein-aether gravitational models are based on the method of the critical
point analysis, where special solutions are determined at the critical points
which describe asymptotic behavior for the evolution of the field equations
\cite{ea01,ea04,ea06,col,ea07}. The plan of the paper is as follows:

In Section \ref{sec2} we briefly present the field equations in
Einstein-aether gravity. Section \ref{sec3} includes the main material of our
study where we present the solution of the field equations for the case where
the aether field is either parallel to the comoving observer (Class A), or tilted (Class B). For these
two classes  we find the relations between the coupling
constants $c_i$ of the theory in order that the field equations do admit solutions.
For both Classes there exist anisotropic solutions, whereas in the first class, where the aether field is parallel to the
comoving observer, there exist an isotropic solution which corresponds
to a FLRW spacetime with nonzero spatial curvature.  Furthermore, the physical
description of the corresponding energy momentum tensor is calculated in all
cases and then used to understand the physical properties of the resulting spacetimes.
Finally, at Section \ref{sec4} we draw our conclusions.

\section{Einstein-\ae ther gravity}

\label{sec2}

The action integral in Einstein-\ae ther gravity is%
\begin{equation}
S_{AE}=\int d^{4}x\sqrt{-g}R+\int d^{4}x\sqrt{-g}\left(  K^{\alpha\beta\mu\nu
}u_{\mu;\alpha}u_{\nu;\beta}+\lambda\left(  u^{\alpha}u_{\alpha}+1\right)  \right)  ,
\label{ai.01}%
\end{equation}
where $u^{\alpha}$ is the aether field i.e. a unit time-like vector field, which
defines the preferred frame. The Lagrange multiplier $\lambda$ ensures the
unitarity of the aether field, $u^{\alpha}u_{\alpha}=-1$. Finally, the Lorentz
violetion terms are introduced by the tensor $K^{\alpha\beta\mu\nu}$ defined
as%
\begin{equation}
K^{\alpha\beta\mu\nu}\equiv c_{1}g^{\alpha\beta}g^{\mu\nu}+c_{2}g^{\alpha\mu
}g^{\beta\nu}+c_{3}g^{\alpha\nu}g^{\beta\mu}+c_{4}g^{\mu\nu}u^{\alpha}%
u^{\beta}. \label{ai.02}%
\end{equation}
where $c_{1}$,~$c_{2}$,~$c_{3}$ and $c_{4}$ are the dimensionless coupling
parameters of the aether field with the gravity.

In \cite{esf1} the first two terms of action integral (\ref{ai.01}) have been
written with the use of the kinematic quantities of the aether field in $1+3$
analysis as follows%
\begin{equation}
S_{EA}=\int\sqrt{-g}dx^{4}\left(  R+c_{\theta}\theta^{2}+c_{\sigma}\sigma
^{2}+c_{\omega}\omega^{2}+c_{\alpha}\alpha^{2} +\lambda\left(  u^{\alpha}u_{\alpha}+1\right)\right)  \label{ae.09}%
\end{equation}
where
\begin{align}
\alpha_{\mu}  &  =u_{\mu;\nu}u^{\nu}~,~\theta=\frac{1}{3}u_{\mu;\nu}h^{\mu\nu
}~,~\sigma_{\mu\nu}=u_{\left(  \alpha;\beta\right)  }h_{\mu}^{\alpha}h_{\nu
}^{\beta}-\theta h_{\mu\nu}~,~\label{ae.08}\\
\omega_{\mu\nu}  &  =u_{\left[  \mu;\nu\right]  }h^{\mu\nu}h_{\mu\nu
}~~,~\sigma^{2}=\sigma^{\mu\nu}\sigma_{\mu\nu}~,~\omega^{2}=\omega^{\mu\nu
}\omega_{\mu\nu},
\end{align}
with $\alpha^{\mu}$ the acceleration, $\theta$ the expansion
rate, $\sigma_{\mu\nu}$ the shear tensor, $\omega_{\mu\nu}$ the
vorticity tensor and $h_{\mu\nu}$ the projective tensor defined as
$h_{\mu\nu}= g_{\mu\nu}-u_{\mu}u_{\nu}$.  The coefficient constants
$c_{\theta},~c_{\sigma},~c_{\omega}$ and $c_{a}$ are related with $c_{1}%
$,~$c_{2}$,~$c_{3}$ and $c_{4}$ as follows%
\begin{equation}
c_{\theta}=\frac{1}{3}\left(  3c_{2}+c_{1}+c_{3}\right)  ~,\ c_{\sigma}%
=c_{1}+c_{3}~,\ c_{\omega}=c_{1}-c_{3}~,\ c_{a}=c_{4}-c_{1}. \label{ae.10}%
\end{equation}

The gravitational field equations of the Einstein-aether theory are obtained
by variation of \eqref{ai.01} with respect to the metric $g^{\mu\nu}$, fielding

\begin{equation}
{G_{\mu\nu}}=T_{\mu\nu}^{\ae }, \label{ai.03}%
\end{equation}
where $G_{\mu\nu}$ is the Einstein tensor and $T_{\mu\nu}^{\ae }$ is the
effective \ae ther energy-momentum tensor which defined as
\begin{align}
{T_{\mu\nu}^{\ae }}  &  =\frac{1}{2}g_{\mu\nu}K^{\alpha\beta\rho\sigma}%
u_{\rho;\alpha}u_{\sigma;\beta}+\frac{1}{2}g_{\mu\nu}\lambda(g^{\alpha\beta
}u_{\alpha}u_{\beta}+1)+\label{en-mo}\\
&  -c_{1}g^{\alpha\beta}(u_{\alpha;\mu}u_{\beta;\nu}+u_{\mu;\alpha}%
u_{\nu;\beta})-c_{2}g^{\alpha\beta}(u_{\mu;\nu}u_{\alpha;\beta}+u_{\beta
;\alpha}u_{\nu;\mu})-c_{3}g^{\alpha\beta}(u_{\alpha;\mu}u_{\nu;\beta}%
+u_{\mu;\alpha}u_{\beta;\nu})+\nonumber\\
&  -c_{4}u^{\alpha}u^{\beta}u_{\mu;\alpha}u_{\nu;\beta}-c_{4}(g^{\lambda
\alpha}u_{\mu}u^{\beta}u_{\lambda;\nu}u_{\alpha;\beta}+g^{\beta\lambda
}u^{\alpha}u_{\nu}u_{\beta;\alpha}u_{\lambda;\mu})-\lambda u_{\mu}u_{\nu
}+\nonumber\\
&  +(\frac{1}{2}K^{\alpha\beta\lambda\kappa}u^{\rho}u_{\kappa;\beta}g_{\rho
\mu}g_{\lambda\nu}+\frac{1}{2}K^{\alpha\beta\lambda\kappa}u^{\rho}%
u_{\kappa;\beta}g_{\rho\nu}g_{\lambda\mu})_{;\alpha}+(\frac{1}{2}%
K^{\alpha\beta\lambda\kappa}u^{\rho}u_{\kappa;\beta}g_{\rho\mu}g_{\alpha\nu
}+\dfrac{1}{2}K^{\alpha\beta\lambda\kappa}u^{\rho}u_{\kappa;\beta}g_{\rho\nu
}g_{\alpha\mu})_{;\lambda}+\nonumber\\
&  -(\frac{1}{2}K^{\alpha\beta\lambda\kappa}u^{\rho}u_{\kappa;\beta}%
g_{\lambda\mu}g_{\alpha\nu}+\frac{1}{2}K^{\alpha\beta\lambda\kappa}u^{\rho
}u_{\kappa;\beta}g_{\lambda\nu}g_{\alpha\mu})_{;\rho}.\nonumber
\end{align}

The equation of motion for the aether field follows by varying $u_{a}$
\begin{equation}
c_{4}g^{\mu\nu}u^{\alpha}u_{\nu;\beta}u_{\mu;\alpha}g^{\kappa\beta}%
-c_{4}g^{\mu\kappa}g^{\alpha\lambda}u_{\lambda;\beta}u^{\beta}u_{\mu;\alpha
}-c_{4}g^{\mu\kappa}u^{\alpha}u_{;\beta}^{\beta}u_{\mu;\alpha}-K^{\alpha
\beta\mu\kappa}u_{\mu;\alpha;\beta}+\lambda g^{\alpha\kappa}u_{\alpha}=0
\label{ai.04}%
\end{equation}
while its constraining unit condition is obtained by varying $\lambda$
\begin{equation}
{u^{a}}{u_{a}+1=0.} \label{ai.05}%
\end{equation}

The above equations make it clear that Einstein-\ae ther gravity is a second-order theory; in a
four dimensional manifold this system comprises of fifteen equations. It is
important to mention that in our consideration we have not assumed any matter content and the 
effective \ae ther energy-momentum tensor is produced from the second term of the action \eqref{ai.01}.

\subsection{Physical Interpretation}

In order to assign a possible physical meaning to the solutions obtained we
calculate, on mass-shell, the Einstein tensor $G_{\mu\nu}=R_{\mu\nu}-\frac
{1}{2}Rg_{\mu\nu}$ and interpret it as an effective energy-momentum tensor of
a perfect fluid by writing $G_{\mu\nu}=T_{\mu\nu}^{(eff)}$. The energy
momentum tensor in the 1+3 decomposition can be written as follows
\begin{equation}
T_{\mu\nu}^{(eff)}=(\rho+p)u_{\mu}u_{\nu}+pg_{\mu\nu}+2q_{(\mu}u_{\nu)}%
+\pi_{\mu\nu},
\end{equation}
where $\rho$ is the energy density of the fluid, $u_{\mu}$ the 4-velosity,
$q_{\mu}$ the heat flux vector, p the pressure and $\pi_{\mu\nu}$ the
anisotropic stress tensor. The relations that make the identification possible
are
\begin{equation}
\Pi_{\mu\nu}=G_{\alpha\beta}h^{\alpha}{}_{\mu}{}h^{\beta}{}_{\nu}= p \,h_{\mu\nu}%
+\pi_{\mu\nu},\quad\pi_{\mu\nu}=\Pi_{\mu\nu}-\frac{1}{3}\Pi^{\kappa}{}_{\kappa
}h_{\mu\nu}=\Pi_{\mu\nu}-p \,h_{\mu\nu}%
\end{equation}%
\begin{equation}
\rho=G_{\mu\nu}u^{\mu}u^{\nu},\quad p=\frac{1}{3}\Pi^{\kappa}{}_{\kappa}%
\end{equation}%
\begin{equation}
q_{\nu}=-G_{\alpha\beta}u^{\alpha}h^{\beta}{}_{n},
\end{equation}
in which $h_{\mu\nu}$ is the projection tensor orthogonal to velosity $u^{\mu
}$ defined by
\begin{equation}
h_{\mu\nu}=g_{\mu\nu}+u_{\mu}u_{\nu},\quad\text{with}\quad u_{\mu}u^{\nu}=-1
\end{equation}


In the next section we begin our analysis by selecting the underlying geometry to be that of a  Bianchi Type $V$ space-time; in this case the field equations
are reduced to ordinary, coupled differential equations with time as the independent variable.

\section{Solution space}

\label{sec3}

The general diagonal Bianchi Type V  line element is
\begin{equation}
ds^{2}=-M(t)^{2}dt^{2}+a(t)^{2}b(t)^{2}dx^{2}+e^{-2x}\left(  a(t)^{4}%
dy^{2}+b\left(  t\right)  ^{4}dz^{2}\right)  . \label{ai.06}%
\end{equation}

If we apply the isometries of the Bianchi V spacetime 
\bal
\xi_{1}=\frac{\partial}{\partial y}, \quad	\xi_{2}=\frac{\partial}{\partial z}, \quad
\xi_{3} =\frac{\partial}{\partial x}+y \frac{\partial}{\partial y}+z \frac{\partial}{\partial z}
\eal
to the \ae ther vector field, also demand that the corresponding
one form be curl-free
and finally utilize \eqref{ai.05} we arrive at %
\begin{equation}
u^{a}=\dfrac{u_{0}\left(  t\right)  }{-M^{2}}\delta_{t}^{a}+\frac
{\text{$\lambda$}_{1}}{a(t)^{2}b(t)^{2}}\delta_{x}^{a},\qquad u_{0}\left(
t\right)  =\sqrt{a(t)^{2}b(t)^{2}+\text{$\lambda$}_{1}^{2}}. \label{ai.08}%
\end{equation}

We choose the time so that $M(t)=a(t)b(t)$; thus, the second component of the equation \eqref{ai.04} assumes the
form
\begin{equation}
\text{$\lambda$}_{1}a^{-6}b^{-6}\left(  (c_{1}+c_{3})\left(  -b^{2}a^{\prime
2}-ab\left(  ba^{\prime\prime}+4a^{\prime}b^{\prime}\right)  +a^{2}\left(
-bb^{\prime\prime}-b^{\prime2}+2b^{2}\right)  \right)  +a(t)^{4}b^{4}%
\lambda\right)  , \label{ai.09}%
\end{equation}
implying that we have to separately investigate the two classes of solutions
where: Class A when $\lambda_{1}=0$ and Class B when $\lambda_{1}\neq0$.

The general approach we adopt in order to reveal the solution space is to
algebraically solve two of the equations in terms of the accelerations
$a^{\prime\prime}(t),b^{\prime\prime}(t)$ and substitute the result into the
rest. In doing so some particular branches appear when the denominators of the
corresponding expressions vanish. This may happen either for specific value
ranges of the constants or for particular relations among $a(t),b(t)$. In what
follows we present all the cases that emerge.

\subsection{Class A solutions}

In a  way similar to \cite{roum} the existence of the solution of the field
equations is directly related to the ranges of values of the coefficients $c_{1},c_{2},c_{3}$ and $c_{4}$ of the theory. In this Class
there are three possible cases of study, corresponding to
\bal
c_{1}+3c_{2}+c_{3}-2 	&=0		\qquad				\tag{Case $A_1$}\\
c_{1}+3c_{2}+c_{3}-2	&\neq 0		\qquad q =0			\tag{Case $A_2$}\\
c_{1}+3c_{2}+c_{3}-2	&\neq 0		\qquad q \neq 0		\tag{Case $A_3$}
\eal
where $q=\frac{1}{5} (5c_1+9c_2+5c_3-4)$. The only non-zero component of \eqref{ai.04} fixes the Lagrange multiplier to

$\lambda=\frac{1}{5 a(t)^4 b(t)^4}\, (a(t) b(t) \left(2 (6 c_2+5 q+4) a'(t) b'(t)-15 c_2 a(t) b''(t)\right)-5 b(t)^2 \left(3 c_2 a(t) a''(t)+(3 c_2-5 q-4) a'(t)^2\right)+5 a(t)^2 (-3 c_2+5 q+4) b'(t)^2)$

which is then substituted into \eqref{ai.03}, giving the final set of equations to be solved in this class.
\subsubsection{Case $A_1$}

The assumpton leads to $q=\frac{6 (1-\text{c2})}{5}$ and substituting in \eqref{ai.03} we calculate the difference $(0,0)-(1,1)=-4$. So there is no solution in this case.

\subsubsection{Case $A_2$}

The assumptions of this case make the $(0,0)$ component of   \eqref{ai.03} %
\begin{equation}
\frac{12}{5}\left(  1-c_{2}\right)  \frac{a^{\prime}b^{\prime}}{ab}-1=0,
\label{ai.13}
\end{equation}

indicating that $c_2\neq 1$ must hold for solutions to exist. As we can easily see the above equation admits the scaling symmetry
$a\rightarrow\omega_{1}a,\quad b\rightarrow\omega_{2}b$; thus, if we make the
replacement
\begin{equation}
a(t)=\exp\left[  \int\left(  a_{1}\left(  t\right)  +b_{1}\left(  t\right)
\right)  dt\right]  ,\quad b(t)=\exp\left[  \int\left(  a_{1}\left(  t\right)
-b_{1}\left(  t\right)  \right)  dt\right]  , \label{rep.01}%
\end{equation}
the aforementioned equation transforms to
\begin{equation}
\frac{12}{5}(1-c_{2})\left(  a_{1}(t)^{2}-b_{1}(t)^{2}\right)  -1=0,
\label{ai.14}%
\end{equation}
which, being a quadratic form in $a_{1}(t),b_{1}(t)$, can be parametrized by%
\begin{equation}
a_{1}(t)=\frac{m}{2}\sinh(f(t)), \,b_1(t)=\frac{m}{2}\cosh(f(t)),  \quad m=\sqrt{\frac{5}{3(c_{2}-1)}}
\end{equation}

If we substitute the above forms of $a_{1}(t), b_{1}(t)$ into the rest
of \eqref{ai.03} we find that a solution exists if $f(t)$ satisfies the
differential equation%
\begin{equation}
2\cosh f(t) \left( 2 m \cosh f(t) + f'(t) \right) =0,\label{ai.16}%
\end{equation}
with solutions
\begin{equation}
f(t)=\frac{i \pi}{2}\left( 2\kappa+1 \right), \, \kappa \in \mathbb{Z},\quad \text{or} \quad
f(t)=-2\tanh^{-1}\left(  \tan\left(  m\left(  t-t_0\right)  \right) \right)     ,\label{ai.17}%
\end{equation}

\begin{itemize}
\item
If $f(t)=\frac{i \pi}{2}\left( 2\kappa+1 \right),$ the solution is real if $c_2<1$ and then
$a(t),b(t)$ becomes
\begin{equation}
a(t)= b(t) = \exp \frac{\epsilon  m}{2}t, \quad  \epsilon = \pm 1,
\end{equation}
yielding the line element
\bal
ds^2= e^{2\epsilon  m t} \left( -dt^2 +dx^2 + e^{-2x} dy^2 + e^{-2x} dz^2 \right).
\eal

The Riemann tensor is zero when $c_2=-\frac{2}{3}$, so we obtain the Minkowski
space-time in this case. The corresponding Ricci scalar is
\begin{equation}
R=-\frac{2(3c_2+2)}{c_2-1} \exp \left(-2\epsilon m \, t \right)
\end{equation}
which has a curvature singularity at $t\rightarrow\pm\infty$
depending on the value of  $\epsilon$.  For this solution the
physical components of the energy momentum tensor describe a perfect fluid
with energy density and pressure component as follows
\bal
\rho=\frac{3c_{2}+2}{c_{2}-1}\exp \left(-2\epsilon m \, t \right), \, p=-\frac{1}{3}\rho.
\eal

\item
If $f\left(  t\right)=-2\tanh^{-1}\left(  \tan\left(  m \left(  t-t_0\right)  \right) \right)$, the solution of the field equations is given by the functions
\bal
a(t) =\sqrt{\sin\left(  m\left(t-t_{0}\right)  \right)  + \cos\left( m\left(t-t_{0}\right)  \right)  },\, 
b(t) =\sqrt{\sin\left(  m\left(t-t_{0}\right)  \right)  - \cos\left(  m\left(t-t_{0}\right)  \right)  }, \label{ai.23}
\eal

Hence, the line element reads
\bal
ds^2 = \cos(2m(t-t_{0})) \left( dt^2 -  dx^2 \right)+e^{-2x} \left( 1+\sin(2m(t-t_{0})) \right) dy^2 - e^{-2x} \left( 1-\sin(2m(t-t_{0})) \right) dz^2
\label{ai.25}
\eal
with $c_2>1$.

The above line element corresponds to a perfect fluid solution with
\begin{align}
\rho &  =\frac{3-m^{2}+3\left(  1+m^{2}\right)  \cos\left(  4m\left(
t-t_{0}\right)  \right)  }{2\cos^{3}\left(  m\left(  t-t_{0}\right)  \right)
}~\\
p  &  =-\frac{1+5m^{2}+\left(  1+m^{2}\right)  \cos\left(  4m\left(
t-t_{0}\right)  \right)  }{2\cos^{3}\left(  m\left(  t-t_{0}\right)  \right)
}%
\end{align}
with equation of state parameter
\begin{equation}
w=-\frac{1+5m^{2}+\left(  1+m^{2}\right)  \cos\left(  4m\left(  t-t_{0}%
\right)  \right)  }{3-m^{2}+3\left(  1+m^{2}\right)  \cos\left(  4m\left(
t-t_{0}\right)  \right)  }.
\end{equation}

It is interesting to note that as the argument $\tau\equiv\left( m\left(  t-t_{0}\right)  \right)  $ spans the prime
interval $[0,2\pi]$ the role of the coordinates $t,x$ as time-like and
space-like respectevely is interchanged (i.e. when $\frac{\pi}{2}<\tau
<\frac{3\pi}{2}$ $t$ is time-like while when $\frac{3\pi}{2}<\tau<\frac{\pi
}{2}$ $x$ is time-like). The variables $y,z$ remain space-like in this interval.

The corresponding Ricci scalar is $\left(  3 \left(  m^{2}+1\right)  \cos(4 m
(t-\text{to}))+7 m^{2}+3\right)  \sec^{3}(2 m (t-\text{to}))$ which 
develops  curvature singularities at finite intervals.
\end{itemize}

\subsubsection{Case $A_3$}

In this case we first solve the (2,2) and (3,3) components of \eqref{ai.03} in terms of the
accelerations $a^{\prime\prime}(t),b^{\prime\prime}(t)$. The denominator has
the term $6\left(  c_2-1\right)+5q$, so we have to check what happens if  $q=\frac{6}{5}(1-c_{2})$; in this case from the components (0,0) and (1,1)  of \eqref{ai.03} we have $(0,0)-(1,1)=-4$, thus we run into incompatibility. 

With $q\neq\frac{6}{5}(1-c_{2})$ we substitute the above mentioned accelerations into
the equation \eqref{ai.03}. The (0,0) component of the
latter equation admits a scaling symmetry $a\rightarrow\omega_{1}a,\quad
b\rightarrow\omega_{2}b$; thus, if we make a replacement similar to ~(\ref{rep.01}), namely 

$a(t)=\exp\left[  \int\left(  -a_{1}\left(  t\right)  +b_{1}\left(  t\right)
\right)  dt\right]  ,\quad b(t)=\exp\left[  \int\left(  a_{1}\left(  t\right)
+b_{1}\left(  t\right)  \right)  dt\right] $, the
(0,0) component of the field equations is simplified as
\begin{equation}
\frac{1}{30}a_{1}(t)^{2}(8(9c_{2}-5q-9))-\frac{1}{30}b\,_{1}(t)^{2}%
(12(6c_{2}+5q-6))=1, \label{ai.26}%
\end{equation}
In order to parametrize the above equation, we have to check the cases when $q$ equals $\frac{1}{5}(9c_{2}-9)$ or not. 

\begin{itemize}
\item
If $q=\frac{9}{5}(c_{2}-1)$ then the equation \eqref{ai.26} gives
\begin{equation}
b_{1}(t)=\pm\left(  \sqrt{6}\sqrt{1-c_{2}}\right)  ^{-1}
\end{equation}

Then we substitute the above form of $a(t), b(t)$ into the equations of the accelerations and we arrive at a
differential equation for $a_{1}(t)$
\begin{equation}
a_{1}^{\prime}(t)\pm4\left(  \sqrt{6}\sqrt{1-c_{2}}\right)  ^{-1}%
a_{1}(t)=0,\label{ai.27}%
\end{equation}
which has the solution $a_1(t)=m_{1}\exp\left(  \mp 4 n t\right)  $ with $m_1$ an integration constant and $n^{-1}=\sqrt{6(1-c_2)}$.

Finally the scale factors $a(t), b(t)$ are given by
\begin{equation}
a(t) = \exp \epsilon \left(2 n t+c\, e^{-4 n t}  \right), \, b(t) = \exp \epsilon \left(2 n t - c\, e^{-4 n t}  \right),\label{ai.28}
\end{equation}
where $\epsilon^{2}=1$, $c$ is a redefinition of $m_1$ and  $c2<1$. The corresponding Ricci scalar reads 
\bal
R=2e^{-4 n t (\epsilon +2)} \left( 64 c^2 n^2 - 3 e^{8 n t}(1-4n^2) \right),
\eal
which has  curvature singularities at $t\rightarrow \pm \infty$ for $\epsilon = -1$ and at $t\rightarrow -\infty$ for $\epsilon = 1$.

The above solution describes a perfect fluid when $\epsilon=1$, and a non perfect one for $\epsilon=-1.$

The energy density and the pressure component are given by
\bal
\rho = e^{-4n(\epsilon+2)t}\left( 3 e^{8nt}(4n^2-1) -64 c^2 n^2 \right), \,
p = -e^{-4n(\epsilon+2)t}\left( e^{8nt}(4n^2-1) +64 c^2 n^2 \right), 
\eal
while the nonzero components of the stress tensor $\pi_{\mu\nu}$ are
\bal
\pi_{yy} = -\pi_{zz} = -32 c n^2 (\epsilon -1) \exp \left( -4nt -2x+4 c \epsilon e^{-4nt} \sqrt{2n^2+1} \right)
\eal

\item
If $(9c_{2}-5q-9)\neq0$ equation \eqref{ai.26} can be parametrized as
\bal
a_{1}(t) =\frac{\sqrt{15}\cosh(f(t))}{2\sqrt{(9c_{2}-5q-9)}}, \, b_{1}(t) =\frac{\sqrt{5}\sinh(f(t))}{\sqrt{2(6c_{2}+5q-6)}}. \label{ai.31} \\
\eal
Now if we substitute the above values of $a_{1}(t),b_{1}(t)$ into the rest of
\eqref{ai.03} we find that a solution exists if $f(t)$ satisfies the
differential equation
\begin{equation}
\cosh(f(t))\left(  \sqrt{10}(6c_{2}+5q-6)f^{\prime}(t)+20\sqrt{6c_{2}+5q-6}\cosh(f(t))\right)  =0, \label{ai.33}
\end{equation}

\begin{itemize}
\item
When $\cosh(f(t))=0$ we have that
\bal
f(t)=\frac{i \pi}{2}\left( 2\kappa+1 \right), \, \kappa \in \mathbb{Z}
\eal

For these functions $f\left(  t\right) $, the solution of the field
equations is determined to be
\bal
a(t)= b(t) = e^{\epsilon k t}, \quad k=\sqrt{\frac{5}{2(6-5q-c_2)}}, \label{ai.34}
\eal
and the line element becomes
\bal
ds^2= e^{4\epsilon k t}\left( -dt^2 +dx^2 +e^{-2x} dy^2 +e^{-2x} dz^2 \right).  \label{ai.36}%
\eal

In the special case when $q=-\frac{1}{5}(6c_{2}+4)$ the Riemann tensor is 0,
and we obtain the Minkowskin space-time.

For any other value of $q$ the corresponding Ricci scalar is
\begin{equation}
R=-6 e^{-4\epsilon k t} \left( 4k^2 - 1 \right),
\end{equation}
which has a curvature singularity at $t\rightarrow\pm\infty$
depending on $\epsilon$, being plus or minus $1$. The energy momentum tensor
describes a perfect fluid, similarly to the previous case. 

\item
When $\cosh f(t) \neq 0$ the solution of the differential equation \eqref{ai.33} is
\begin{equation}
f(t)=-2\tanh^{-1}  \left(\tan\left(  A \left(  t-t_0 \right)  \right) \right), \quad A=\sqrt{10(6c_{2}+5q-6)^{-1}}  , \label{ai.39}
\end{equation}
which implies that,
\bal
a(t) = \cos^{1/4}2A(t-t_0) \left( \frac{1+\tan A(t-t_0)}{1-\tan A(t-t_0)} \right)^{-B}, \,
b(t) = \cos^{1/4}2A(t-t_0) \left( \frac{1+\tan A(t-t_0)}{1-\tan A(t-t_0)} \right)^B
\eal
with $B=\sqrt{3(8(6-5A^2q))^{-1}}$.

The corresponding line element is
\bal
ds^2 = \cos(2At)\left(  -dt^{2}+dx^{2} + e^{-2x} \left(  \left( \frac{1+\tan A t}{1-\tan A t} \right)^{-4B} dy^2  
+\left( \frac{1+\tan A t}{1-\tan A t} \right)^{4B} dz^2 \right) \right),
\eal
where we removed $t_0$ by a redefinition of the coordinate $t \mapsto t+t_0$. It is easy  to see that this line element describes a perfect fluid solution.

The Ricci scalar is given by
\begin{equation}
R=-\left( 3 + A^2(9-32B^2)+ 3(A^2+1) \cos(4 At) \right) \sec^3(2A t),
\end{equation}
which diverges for $t = \frac{\pi}{4A}(2\kappa+1), \, \kappa \in \mathbb{Z}$.
\end{itemize}
\end{itemize}

We proceed our analysis with the derivation of the second-class of solutions
in which $u^{a}$ is not the comoving observer.

\subsection{Class B solutions}

For the second class of solutions where $\lambda_{1}\neq0,$ the strategy is
now to first solve the second component of (\ref{ai.04}) for the Lagrange
multiplier $\lambda(t)$ obtaining:
\begin{equation}
\lambda(t)=-(c_{1}+c_{3})\left(  -ab^{2}a^{\prime\prime}-4aba^{\prime
}b^{\prime2}a^{\prime2}-a^{2}bb^{\prime\prime2}b^{\prime2}+2a^{2}b^{2}\right)
\left(  ab\right)  ^{-4}. \label{ai.42}%
\end{equation}

Next we substitute this $\lambda(t)$ into the first component of \eqref{ai.04}
and also into the components of \eqref{ai.03}.

If, from the resulting equations, we solve the first component of
\eqref{ai.04} and the (2,2) component of \eqref{ai.03} in terms of the
accelerations $a^{\prime\prime}(t),b^{\prime\prime}(t)$, the possible
vanishing of the following expression%

\[
(a^{2}b^{2}(c_{1}+c_{3}+1)+\text{$\lambda$}_{1}^{2}(c_{1}+c_{3}))\,(a(t)^{2}%
b(t)^{2}(c_{1}+3c_{2}+c_{3})+2c_{2}\text{$\lambda$}_{1}^{2})
\]

appearing in the denominators of $a^{\prime\prime}(t),b^{\prime\prime}(t)$
must be considered, giving rise to the following cases

\bal
a^2 b^2 (c_1+c_3+1)+\lambda_1^2(c_1+c_3)=0	&	& c_1+c_3+1 \neq0						\tag{Case $B_1$}\\
c_2=0											&	& c_1+c_3=0							\tag{Case $B_2$}\\
a^2 b^2(c_1+3c_2+c_3)+2c_2 \lambda_1^2=0 	&    & c_1+3c_2+c_3\neq0,	\,q \neq 0		\tag{Case $B_3$}\\
c_1+c_3\neq0									&	&										\tag{Case $B_4$}
\eal

\subsubsection{Case $B_1$}

This case finally results into $a,b$ being both constants functions. When we substitute these constant values in \eqref{ai.03}, we end up
with $(0,0)+(1,1)=-2(c_{1}+c_{3}+1)=0$ which is impossible; thus there is no solution in this case.

\subsubsection{Case $B_2$}

With the above assumption the constrain equation, i.e. the (0,0) 
component of \eqref{ai.03}, becomes
\begin{equation}
\frac{8a^{\prime}b^{\prime}}{ab}+\frac{2a^{\prime2}}{a^{2}}+\frac{2b^{\prime
2}}{b^{2}}-3=0.
\end{equation}

As before, the existing scaling symmetry indicates that the replacement%

\begin{equation}
a(t)=\exp\left[  \frac{1}{2}\int\left(  \sqrt{3}\sinh(f(t))\,+\cosh(f(t))\right)  dt\right]  ,
b(t)=\exp\left[ \frac{1}{2}\int\left(  \cosh(f(t))-\sqrt{3}\sinh(f(t))\right)  dt\right]  , \label{ai.44}
\end{equation}
satisfies the above expression, while the rest of the equations \eqref{ai.03}
are solved if $f(t)$ obeys the first order differential equation

\begin{equation}
-2\sinh(f(t))\left(  f^{\prime}(t)+2\sinh(f(t))\right)  =0. \label{ai.45}%
\end{equation}

Its solutions read
\bal
f(t) = i\kappa \pi, \, \kappa \in \mathbb{Z}, \quad \text{or} \quad f(t)=2\coth^{-1}  e^{2t-m_{1}}  \label{ai.48},
\eal
where $m_1$ is an integration constant.

\begin{itemize}
\item
If $f(t) = i\kappa \pi$ the solution is
\begin{equation}
a(t)=e^{\frac{t\epsilon}{2}},\quad b(t)=e^{\frac{t\epsilon}{2}},
\label{ai.49}
\end{equation}
and the line element becomes
\begin{equation}
ds^2 = e^{2\epsilon t} \left( -dt^2 +dx^2 +e^{-2x} dy^2 +e^{-2x} dz^2 \right), \label{ai.50}
\end{equation}
with the Riemann tensor equal to zero, indicating the Minkowski space time.

\item
If $ f(t)=2\coth^{-1}  e^{2t-m_{1}}$ the  solution is:
\bal
a(t)=e^{-\frac{t}{2}}\left(  \frac{e^{m_{1}}-e^{2t}}{e^{m_{1}}+e^{2t}}\right)^{\frac{\sqrt{3}}{4}}\sqrt[4]{1-e^{4t-2m_{1}}}, \quad
b(t)=e^{-\frac{t}{2}}\left(  \frac{e^{m_{1}}-e^{2t}}{e^{m_{1}}+e^{2t}}\right)^{-\frac{\sqrt{3}}{4}}\sqrt[4]{1-e^{4t-2m_{1}}},\label{ai.52}
\eal
with the corresponding line-element
\begin{equation}
ds^{2}=\kappa^2 \sinh(2\tau) \left( d\tau^2 - dx^2 +e^{-2x} \tanh^{\sqrt{3}}(\tau) dy^2 +e^{-2x} \tanh^{-\sqrt{3}}(\tau) dz^2 \right),
\end{equation}
where $\kappa^2=2e^{-m1}$ and $\tau=t+ln(\frac{\kappa}{\sqrt{2}})$. 

The Ricci tensor is derived to be $R_{\mu\nu}=0$; thus the above line element describes a vacuum
solution. The Kretsmann scalar equals
\bal 
Kr\equiv R^{\mu\nu\kappa\lambda}R_{\mu\nu\kappa\lambda}=\frac{96}{\kappa^4}\,\text{csch}^6(2\tau),
\eal 
indicating the existence of a singularity at $\tau=0$. Note
that $\kappa$ is a gravitational essential constant and that the time--like character between $\tau,x$ is interchanged as $\tau$ spans the interval $(-\infty,\infty)$.
\end{itemize}

\subsubsection{Case $B_3$}

In this case we end up with $a(t) b(t)=\lambda_1\sqrt{-2c_2}(c_1+3c_2+c_3)^{-1/2}$ and if we substitute this expression into the components of
\eqref{ai.03} we can see that $a(t)$ must be constant, which entails the result
$(0,0)+(1,1)=-2(c_{1}+c_{3}+1)=0$. However, this relationship makes the rest
of equation \eqref{ai.03} incompatible with the assumptions of the given
branch; we thus conclude that there is no solution.

The above analysis corresponds to the last case where the denominators vanish. Thus we can now use
the forms of the accelerations $a^{\prime\prime}(t),b^{\prime\prime}(t)$ and
replace them into equation \eqref{ai.03}. This action results in the (0,1) component
being
\begin{equation}
2\text{$\lambda_{1}$}(c_{1}+c_{3})\left(  \lambda_{1}ba^{\prime}+a\left(
b\sqrt{a^{2}b^{2}+\lambda_{1}^{2}}+\lambda_{1}b^{\prime}\right)  \right)
\left(  ab\right)  ^{-3}=0.\label{ai.53}%
\end{equation}
Various cases will emerge from the above equation. When $c_{1}+c_{3}=0$, the
only independent components of \eqref{ai.03} are the (0,0) and (1,1) which are
quadratic in the velocities. If we differentiate these with respect to $t$,
use again the accelerations and subtract appropriate linear combinations of
the expressions themselves, we end up with the following two cases:

\begin{itemize}
\item
If $c_2\neq 0, \, c_1+c_3=0, \, a b=c_b$ then the combination (0,0)-(1,1) of the components of \eqref{ai.03} reads $-\frac{4 \text{c2} \lambda _1^2}{\text{cb}^2}-4=0$ implying $\text{cb}=i \sqrt{\text{c2}} \lambda _1$ and thus a neutral signature. With this value for $\text{cb}$ the (0,0) component of \eqref{ai.03} becomes
\begin{equation}
-\frac{4 a'(t)^2}{a(t)^2}-1=0, \label{ai.54}
\end{equation}
which has the solution
\begin{equation}
a(t)=e^{m_1+\frac{i t \epsilon }{2}}  \label{ai.55}
\end{equation}
with $\epsilon=\pm1$,  $m_1$ an integration constant,  and  all  equations have now been sattisfied.

At this stage the redefinitions $\left\{y\to \mu  e^{-2 m_1} Y,z\to \frac{e^{2 m_1} Z}{\mu },t\to i T,x\to X,\text{c2}\to \frac{\mu ^2}{\lambda _1^2}\right\}$ cast the line element into the simple final form 
\begin{equation}
ds^{2}=\mu ^2 \left(-dT^2-dX^2+e^{-2 T \epsilon -2 X} dY^2 +e^{2 T \epsilon -2 X} dZ^2 \right). \label{ai.56} %
\end{equation}
while the aether one-form transform into $u_\mu=\left\{-\sqrt{\mu ^2-\text{$\lambda_1 $}^2},\text{$\lambda_1$},0,0\right\}$

which, in order to be real, implies the restriction $\text{$\lambda $1}^2<\left| \mu \right| ^2$. 
 
The above line-element represents a singularity free geometry, on account of having vanishing covariant derivative of its Riemmann tensor and all fourteen curvature scalars monomialls of the Ricci scalar $R=\frac{8}{\mu^2}$. Note that, since the space admits only one extra Killing field $\xi_4=(1,0,\epsilon Y, -\epsilon Z)$, we have the intersting case of a CSI space time \cite{milson}  

Despite the non physical character of a neutral singnature line element it is not unreasonable to  investigate the  resulting effecting $T_{\mu \nu}$ in terms of perfect fluid:%
\begin{equation}
\rho=\frac{2}{\mu ^2}~,~p=-\frac{2}{\mu ^2}~,~q_{\kappa}=0,~\pi_{\mu\nu}=0
\end{equation}
inferring that it corresponds to an isotropic perfect fluid with an equation of state $p+\rho=0$ and zero heat conduction, i.e. it mimics that of a
cosmological constant.

\item
If $c_2\neq0, \,c_1+c_3=0,\, a b\neq c_b$ then with the application of the same steps as before we end up with $a b=\text{const.}$,
which contradicts the initial assumption of this case. Thus there is no solution here although, of course, the previous solution hold true.
\end{itemize}

\subsubsection{Case $B_4$}

Now we can solve equation \eqref{ai.53} as a differential equation for $b(t)$
obtaining
\begin{equation}
b(t)=\frac{2\text{$\lambda$}_{1}^{2}e^{m_{1}+t}}{e^{2t}a(t)-\text{$\lambda$%
}_{1}^{2}e^{2m_{1}}a(t)}, \label{ai.61}%
\end{equation}
with $m_1$ an integration constant.

When this is substituted into \eqref{ai.03} we get a differential equation for
$a(t)$ which leads to the final scale factors
\bal
a(t)=\frac{m_2e^{t/2}}{\sqrt{e^{2t}-\lambda_1^2 e^{2m_1}}}, \,
b(t)=\frac{2\lambda_1^2 e^{m_1+\frac{t}{2}}}{m_2\sqrt{e^{2t}-\lambda_1^2 e^{2m_1}}} \label{ai.63}%
\eal
under the assumpion $c_{1}+3c_{2}+c_{3}=2$.

The resulting line element is 
\bal
ds^2 = \lambda_1^2 \text{csch}^2t \left( -dt^2 +dx^2 +e^{-2x} dy^2 +e^{-2x} dz^2 \right),
\eal
where we have used the transformation $t \mapsto t + m_1 +\ln \lambda_1, \, x \mapsto x + m_1^2 +\ln 2, \, y \mapsto 4m_2^{-2} \lambda_1^2 y, \, z\mapsto e^{-2m_1} \lambda_1^{-2} m_2^2 z$.

Furthermore, it holds that $R_{\lambda\mu\nu\kappa}= \frac{1}{\lambda_{1}^{2}}(g_{\lambda\nu}g_{\mu\kappa}-g_{\lambda\kappa}g_{\mu\nu})$, revealing the
space-time as maximally symmetric with Ricci scalar $R=\frac{12}{\lambda_1^2}.$. If we further investigate the physical properties of the energy-momentum tensor
for the above solution, we find that%
\begin{equation}
\rho=\frac{3}{\lambda_{1}^{2}}~,~p=-\frac{3}{\lambda_{1}^{2}}~,~q_{\kappa}=0,~\pi_{\mu\nu}=0
\end{equation}
inferring that it corresponds to an isotropic perfect fluid with an equation of state $p+\rho=0$ and zero heat conduction, i.e. it mimics that of a
cosmological constant.

\section{Conclusion}

\label{sec4}

In this work we studied the existence of exact solutions for the field
equations in the Lorentz violating theory known  as Einstein-aether gravity.
For the background geometry we assumed that of a Bianchi Type V
spacetime, with a diagonal scale factor matrix dictated by the vanishing of the $G_{tx}$  constraint equation.  
Note that this form of the line-element, although with two independent functions of time, is not Locally Rotationally Symmetric, i.e. it does not admit a fourth Killing field. We have also restricted the form of the aether field by imposing on it the symmetries of the geometry; thus arriving at two possible cases, that of
tilted or non-tilted $u_\mu$. For both models of our study we determined the
field equations and exhibited their solution space. Like in the previous similar work of ours concerning Bianchi Type III geometry \cite{roum}, we also here find that exact solutions exist only for specific values of the coupling parameters of the Einstein-aether theory. 

When the aether field is parallel to the comoving observer, we found that
there exists an exact isotropic solution in case $A_2$, when $c_{1}+3c_{2}+c_{3}-2\neq0,\, q=0$. In this case,
the spacetime reduces to the Friedmann--Lema\^{\i}tre--Robertson--Walker
spacetime with non-zero spatial curvature. On the other hand, in case $A_3$, when 
$c_{1}+3c_{2}+c_{3}-2\neq0,\,q\neq0$ both anisotopic and isotropic solutions are found; the latter case being when
$(9c_{2}-5q-9)\neq0$, an isotropic solution is found.

In a similar way when the aether field is not parallel to the comoving
observer anisotropic and isotropic exact solutions are found for specific
relations between the coupling constants of the Einstein-aether theory. There
are two possible constraining relations for the coupling constants, when exact
solutions exist. These occur in case $B_2$ where $c_2=0,\, c_1+c_3=0,$ and in case $B_6$ where $c_1+c_3\neq0$. 

Lastly, we  would like to briefly comment about the LRS case: The scale factor matrix is now $ \gamma_{\mu \nu}=diag\{a^2,a^2,b^2\} $ which results in a non-zero $G_{tx}$ component of the Einstein tensor. Now the geometry admits the fourth Killing field $\xi_4=-z \frac{\partial}{\partial y}+y \frac{\partial}{\partial z}$. The aether field remains the same, since it commutes with $\xi_4$, and thus the classification according to $\lambda_1$ being zero or non-zero is also valid. There is however a substantial difference in the solution space. For example in the class A case ($\lambda_1=0$) there is only one family of solutions described by the scale factors $a(t)=m_2 \exp{\frac{\left(\sqrt{2} t\right) \epsilon }{m_1\sqrt{(2-Q)}}}$ and $b(t)=m_1 a(t)$ where $\epsilon=\pm 1, \,Q=c_1+3c_2+c3<2$ for the solution to be real, with the Minkowski space-time corresponding to the value $Q=0$.  Note also that, if a free electromagnetic field was considered instead of the aether field, the line element used in the present work would lead only to the vacuum solution whereas the LRS element would have non-trivial solutions. All these issues are under consideration and will be presented in a future publication.

\end{document}